\documentclass[aps,11pt,nofootinbib,preprintnumbers,showpacs]{revtex4}

\usepackage{amssymb, amsfonts, amsmath, bm}
\usepackage{amsfonts}
\usepackage[dvipdfmx]{graphicx,color}

\begin{document}

\title{
Asymptotically flat multi-black lenses 
}
\vspace{2cm}

\author{Shinya Tomizawa${}^{1,2}$\footnote{tomizawasny@'stf.teu.ac.jp} and Taika Okuda${}^{2}$\footnote{c0113109e1@edu.teu.ac.jp}  
} 
\vspace{2cm}
\affiliation{
${}^1$ Department of Liberal Arts, Tokyo University of Technology, 5-23-22, Nishikamata, Otaku, Tokyo, 144-8535, Japan, \\
${}^2$ School of Computer Science, Tokyo University of Technology, 1404-1, Katakuramachi, Hachioji City, Tokyo, 192-0982, Japan
}

\begin{abstract} 
We present an asymptotically flat and stationary multi-black lens solution with bi-axisymmetry of $U(1)\times U(1)$ as a supersymmetric solution in the five-dimensional minimal ungauged supergravity.  We show that the spatial cross section of each degenerate Killing horizon admits different lens space topologies of $L(n,1)=S^3/{\mathbb Z_n}$ as well as a sphere $S^3$. 
Moreover, we show that in contrast to the higher dimensional Majumdar-Papapertrou multi-black hole and multi-BMPV black hole spacetimes,  the metric is smooth on each horizon even if the horizon topology is spherical.  
\end{abstract}
\date{\today}
\maketitle

\section{Introduction}\label{sec:1}

In recent years, in string theory and the various contexts of AdS/CFT correspondence, higher dimensional black holes and other extended black objects have played an important role~\cite{Tangherlini:1963bw,Myers:1986un,Emparan:2001wn,Pomeransky:2006bd,Emparan:2008eg}. 
In particular, physics of asymptotically flat black holes  in the five-dimensional minimal supergravity  (Einstein-Maxwell-Chern-Simons theory) has been the subject of increased attention, as it describes a low-energy limit of string theory.
Various types of black hole solutions in the theory have so far been found, with the help of recent development of solution generating techniques~\cite{Gauntlett:2002nw,Bellorin:2006yr,Bellorin:2007yp,Gutowski:2005id,Gauntlett:1998fz,Grover:2008ih,Gutowski:2007ai,
Reall:2002bh,Breckenridge:1996is,Gibbons:1999uv,Elvang:2004rt,Elvang:2004ds,Elvang:2004xi,Compere:2009iy,Gibbons:2013tqa,Kunduri:2014iga,Kunduri:2013vka,Niehoff:2016gbi,Tomizawa:2012nk,Mizoguchi:2012vg,Mizoguchi:2011zj,Tomizawa:2008qr,Tomizawa:2008rh,Matsuno:2008fn,Tomizawa:2008tj,Nakagawa:2008rm,Tomizawa:2007he,Ishihara:2006pb,Ishihara:2006iv,Ishihara:2005dp,Bena:2004de,Bena:2007kg,Bena:2009qv}.

\medskip
 The topology theorem for stationary black holes generalized to five dimensions~\cite{Cai:2001su,Galloway:2005mf,Hollands:2007aj,Hollands:2010qy} states that the topology of the spatial cross section of the event horizon is restricted to either a sphere $S^3$, a ring $S^1\times S^2$, or lens spaces $L(p,q)$ $(p,q:$ coprime integers), if one assumes that the spacetime is asymptotically flat and admits three commuting  Killing vector fields, a timelike Killing  vector field and two axial Killing vector fields under the dominant energy condition. 
As for the first two cases, one knows the corresponding exact solutions in five-dimensional Einstein theory~\cite{Myers:1986un,
Emparan:2001wn,Pomeransky:2006bd} and minimal ungauded supergravity theory~\cite{Breckenridge:1996is,Elvang:2004rt,Elvang:2004ds}.  On the contrary, a regular black hole solution with a lens space topology, at present, has not been found to the five-dimensional vacuum Einstein equation, although a few authors have attempted to construct such a black lens solution by using the combination of the rod diagram and the inverse scattering method~\cite{Evslin:2008gx,Chen:2008fa}. 

\medskip
Recently, there has been a new development in this field. Asymptotically flat and stationary black lens solutions, whose horizon topology is restricted to $L(2,1)=S^3/{\mathbb Z}_2$, were constructed by Kunduri and Lucietti as supersymmetric solutions to the five-dimensional minimal ungauged supergravity~\cite{Kunduri:2014kja} and $U(1)^3$ supergravity~\cite{Kunduri:2016xbo}. Furthermore, the more general black lenses with the horizon topologies of $L(n,1)=S^3/{\mathbb Z}_n\ (n\ge3)$ were also constructed  by one of the present authors in the former theory~\cite{Tomizawa:2016kjh}. 
The basic strategy to get these supersymmetric black lens solutions is to use the well-known method developed by Gauntlett et al. in~\cite{Gauntlett:2002nw}. 

\medskip
It is well known that the Majumdar-Papapetrou solution to the four dimensional Einstein-Maxwell equation describes an arbitrary number of charged static black holes in an asymptotically flat spacetime by a balance of electromagnetic and gravitational forces~\cite{Majumdar:1947eu,Papapetrou:1948jw}. 
Such an asymptotically flat, static multi-black hole solution was perviously generalized to higher dimensional Einstein-Maxwell theories~ \cite{Myers:1986rx}.  Furthermore,  a multi-black hole solution in a rotational case (this is often called multi-BMPV black hole) was constructed  in five-dimensional minimal supergravity~\cite{Candlish:2009vy}.  As shown in~\cite{Welch:1995dh, Candlish:2007fh, Candlish:2009vy}, in contrast to the four-dimensional  Majumdar-Papapetrou solution~\cite{Hartle:1972ya}, these solutions generalized to higher dimensions do not admit smoothness of the metric at horizons, whereas for the concentric multi-black ring solution~ \cite{Elvang:2004rt}, the spacetime is known to be analytic at each event horizon. Therefore, it is not entirely clear whether there does exist a regular multi-black lens solution in higher dimensions, simply because a black lens solution with a single horizon exists. 

\medskip
The purpose of this paper is to construct an exact solution describing an arbitrary number of charged rotating black lenses with smooth horizons as an asymptotically flat and stationary supersymmetric solution in five-dimensional minimal supergravity.  This work is essentially based on the previous works of~\cite{Kunduri:2014kja,Tomizawa:2016kjh}, where the strategy is to use the Gibbons-Hawking space as a hyper-K\"ahler base space and allow the harmonic functions to have $n$ point sources with appropriate coefficients. We show that by imposing appropriate boundary conditions on the parameters, the $n$ point sources denote degenerate Killing horizons with the topologies of different lens spaces $L(n_i,1)=S^3/Z_{n_i}\ (i=1,\cdots,n)$.
Moreover, introducing an appropriate coordinate system, we also show that the metric and Maxwell's field strength are smooth on each horizon  in contrast to the higher dimensional  Majumdar-Papapetrou solutions and multi-BMPV black hole solution.

\medskip

This paper is organized as follows: 
In Sec.~\ref{sec:solution}, following the work of Gauntlett et al.~\cite{Gauntlett:2002nw}, we present the supersymmetric solution on the Gibbons-Hawking base space, which describes multi-black lenses in the five-dimensional minimal ungauged supergravity.
The solution admits three commuting Killing vector fields, {\it i.e.},  the stationary Killing vector field, two mutually commuting axial Killing vector fields so that the isometry group of the spacetime is ${\mathbb R}\times U(1)\times U(1)$. 
In Sec. III, we impose the boundary conditions so that the spacetime is asymptotically flat, admits no closed timelike curves (CTCs) on and outside the horizons, neither conical nor curvature singularities appear in the domain of outer communications, and no orbifold singularity and  no Dirac-Misner string exists on the axis. 
Section IV is devoted to study some physical properties of such multiple black lenses. 
In Sec. V, we summarize our result and discuss further generalization.

\section{Black lens solution}
\label{sec:solution}

We would like to consider supersymmetric solutions in the five-dimensional minimal ungauged supergravity, whose bosonic Lagrangian is described by  the Einstein-Maxwell-Chern-Simons theory: 
\begin{eqnarray}
\mathcal L=R \star 1 -2 F \wedge \star F -\frac 8{3\sqrt 3}A \wedge F \wedge F \,, 
\end{eqnarray}
where $F=d A$ is the Maxwell field. 
The metric and gauge potential $1$-form are given by
\begin{eqnarray}
\label{metric}
ds^2&=&-f^2(dt+\omega)^2+f^{-1}ds_{M}^2,\\
A&=&\frac{\sqrt 3}{2} \left[f(d t+\omega)-\frac KH(d \psi+\chi)-\xi \right]\,. 
\end{eqnarray}
where we choose the hyper-K\"ahler metric  $ds^2_M$ to be the Gibbons-Hawking metric
\begin{eqnarray}
ds^2_M&=&H^{-1}(d\psi+\chi)^2+Hdx^idx^i, \\
d \chi& =& * d H \,,\label{eq:dchi}
\end{eqnarray}
where $\{x^i\}=(x,y,z)\ (i=1,2,3)$ are Cartesian coordinates on $\mathbb{E}^3$ and 
$\partial/\partial \psi$ is a triholomorphic Killing vector. 
Furthermore, 
\begin{eqnarray}
f^{-1}&=&H^{-1}K^2+L,\\
\omega&=&\omega_\psi(d\psi+\chi)+\hat \omega,\\
\omega_\psi&=&H^{-2}K^3+\frac{3}{2} H^{-1}KL+M, \\ 
*d\hat\omega&=&HdM-MdH+\frac{3}{2}(KdL-LdK),\label{eq:dhatomega}\\
\qquad d \xi &=&-* d K \,.\label{eq:dxi}
\end{eqnarray}
Here $H, K, L, M$ are harmonic functions on $\mathbb E^3$, where it should be noted that there exists a gauge freedom of redefining harmonic functions~\cite{Bena:2005ni} 
\begin{eqnarray}
K\to K+aH,\qquad L\to L-2aK-a^2H,\qquad M\to M-\frac{3}{2}aL+\frac{3}{2}a^2K+\frac{1}{2}a^3H,\label{eq:trans}
\end{eqnarray}
where $a$ is an arbitrary constant. 
In fact, under the transformation (\ref{eq:trans}),  ($f, \omega_\psi, \chi$) remain invariant, whereas the 1-form $\xi$ undergoes a change as
$\xi\to \xi-a\chi$. 
Since this transformation merely amounts to the gauge transformation 
$A\to A+a d \psi$,  the transformation (\ref{eq:trans}) makes the bosonic sector invariant.

\medskip  
Following the papers on supersymmetric black lenses in~\cite{Kunduri:2014kja} and~\cite{Tomizawa:2016kjh}, 
 we consider the next harmonic functions with $n$ point sources
\begin{eqnarray}
H&=&\sum_{i=1}^n\frac{h_i}{r_i}:=\sum_{i=1}^n\frac{n_i}{r_i}, \label{Hdef}\\
M&=&m_0+\sum_{i=1}^n\frac{m_i}{r_i},\label{Mdef}\\
K&=&\sum_{i=1}^n\frac{k_i}{r_i},\label{Kdef}\\
L&=&l_0+\sum_{i=1}^n\frac{l_i}{r_i}.\label{Ldef}
\end{eqnarray}
Here, each $n_i$ takes not only positive but also negative integers ($n_i=\pm1,\pm2,\cdots$), and 
$r_i:=|{\bm r}-{\bm r_i}|=\sqrt{(x-x_i)^2+(y-y_i)^2+(z-z_i)^2}$ with constants $(x_i,y_i,z_i)$.  

From Eq.~(\ref{eq:dchi}), (\ref{eq:dhatomega}) and (\ref{eq:dxi}), the 1-forms ($\chi, \xi, \hat \omega$) are determined as
\begin{eqnarray}
\chi&=&\sum_{i=1}^nh_i\tilde\omega_i,\\
\xi&=-&\sum_{i=1}^nk_i\tilde\omega_i,\\
\hat \omega&=&\sum_{i,j=1(i\not=j)}^n\left(h_im_j+\frac{3}{2}k_i l_j \right)\hat \omega_{ij}-\sum_{i=1}^n\left(m_0h_i+\frac{3}{2}l_0k_i\right)\tilde\omega_i+c,
\end{eqnarray}
where $c$ is  a constant and the 1-forms $\tilde\omega_{i}$ and $\hat \omega_{ij}$ ($i\not=j$) on ${\mathbb E}^3$ are, respectively,
\begin{eqnarray}
\tilde\omega_i&=&\frac{z-z_i}{r_i}\frac{(x-x_i)dy-(y-y_i)dx}{(x-x_i)^2+(y-y_i)^2},\\
\hat\omega_{ij}&=&-\frac{({\bm r}-{\bm r}_i)\cdot ({\bm r}-{\bm r}_j)}{r_ir_j}\frac{\left[ ({\bm r}_i-{\bm r}_j)\times ({\bm r}-\frac{{\bm r_i}+{\bm r_j}}{2})\right]_kdx^k}{\left|({\bm r}_i-{\bm r}_j)\times({\bm r}-\frac{{\bm r_i}+{\bm r_j}}{2})\right|^2}.
\end{eqnarray}
Throughout this paper, we set $x_i=y_i=0$ for all $i$,  by which $x\partial/\partial y-y\partial/\partial x$ becomes another Killing field and assume $z_i<z_j$ for $i<j$.
In this case,  $\tilde \omega_i$ and $\hat \omega_{ij}$ are simply written in spherical  coordinates 
$(x,y,z)=(r\sin\theta\cos\phi, \sin\theta\sin\phi, r\cos\theta)$ as
\begin{eqnarray}
\tilde \omega_i &=&\frac{z-z_i}{r_i}d \phi, \\
\hat\omega_{ij}&=&\frac{r^2-(z_i+z_j)r\cos\theta+z_iz_j}{z_{ji}r_ir_j},
\end{eqnarray}
where $z_{ji}:=z_j-z_i$.

\section{Boundary conditions}\label{sec:boundary}

In the present paper, we would like to obtain a supersymmetric multiple black lens solution such that it describes physically interesting spacetime. We impose suitable boundary conditions at (i) infinity $r\to\infty$, (ii) horizon ${\bm r}={\bm r}_i\ (i=1,\cdots,n)$, and (iii)  on the $z$-axis $x=y=0$  in the Gibbons-Hawking space: 
(i)  At infinity $r\to\infty$, the spacetime is asymptotically flat,  
(ii) each surface ${\bm r}={\bm r}_i\ (i=1,\cdots,n)$ should correspond to a smooth degenerate Killing horizon whose spatial cross section has a topology of the lens space $L(n_i,1)=S^3/{\mathbb Z}_{n_i}$, and   
(iii) on the $z$-axis $x=y=0$ in the Gibbons-Hawking space, we require that there should appear no Dirac-Misner string, and  orbifold singularity must be eliminated. 
Moreover, besides these boundaries, in the domain of outer communication, the spacetime allows neither CTCs nor (conical and curvature) singularities.

\subsection{Infinity}
\label{sec:bc_infinity}
First of all, let us consider the boundary condition to satisfy asymptotic flatness. 
For $r \to \infty$, the metric functions ($f, \omega_\psi$) behave as 
\begin{eqnarray}
f^{-1}&\simeq &l_0+\left[\left(\sum_{i}k_i\right)^2+\sum_{i}l_i\right] r^{-1}, \qquad 
\omega_\psi\simeq m_0+\frac{3}{2}l_0\sum_{i} k_i. 
\end{eqnarray}
Since the 1-forms $\tilde\omega_{i}$ and $\hat\omega_{ij}$ are approximated as
\begin{eqnarray}
\tilde \omega_i \simeq \cos\theta d\phi,\qquad \hat\omega _{ij}\simeq \frac{d\phi}{z_{ji}},
\end{eqnarray}
the 1-forms $\chi$ and $\omega$ behave as, respectively, 
\begin{eqnarray}
\chi&=& \sum_{i} h_i \hat\omega_i\simeq \sum_{i} n_i\cos\theta d\phi,\\
\omega
                    &\simeq & \left(m_0+\frac{3}{2}l_0\sum_{i}k_i\right)\left(d\psi +\cos\theta d\phi \right)-\sum_{i}\left(m_0h_i+\frac{3}{2}l_0k_i\right)\cos\theta d\phi \notag \\
                    &&+\left(\sum_{i,j(i\not =j)}\frac{h_im_j
                    +\frac{3}{2}k_il_j}{z_{ji}} +c\right)d\phi.
\end{eqnarray}
The asymptotic flatness demands that the parameters should satisfy
\begin{eqnarray}
l_0&=&1,\label{eq:l0}\\
\sum_{i=1}^nn_i&=&1, \label{eq:sum-n_i}\\
c&=&-\sum_{i,j(i\not =j)}\frac{h_im_j+\frac{3}{2}k_il_j}{z_{ji}},\label{eq:c}\\
m_0&=&-\frac{3}{2}\sum_{i}k_i.\label{eq:m0}
\end{eqnarray}
In terms of the radial coordinate $\rho=2\sqrt{r}$, 
the metric asymptotically ($\rho \to \infty$) behaves as
\begin{eqnarray}
ds^2&\simeq& -dt^2+d\rho^2+\frac{\rho^2}{4}\left[(d\psi+\cos\theta d\phi)^2+ d\theta^2+\sin^2\theta d\phi^2 \right].
\end{eqnarray}
This coincides with the metric of Minkowski spacetime, where the metric of  $S^3$ 
is written in terms of Euler angles ($\psi, \phi, \theta)$. 
The avoidance of conical singularities requires the range of angles to be 
$0\le \theta \le \pi$, $0\le \phi <2\pi $ and $0\le \psi <4\pi$ with the identification
$\phi\sim \phi+2\pi$ and $\psi\sim \psi+4\pi$.

\subsection{Horizon}
Next, we show that each point source ${\bm r}={\bm r}_i\ (i=1,\cdots,n)$ denotes a degenerate Killing horizon whose spatial topology is a lens space $L(n_i,1)=S^3/\mathbb{Z}_{n_i}$. 
In terms of the radial coordinate redefined by $r:=|{\bm r}-{\bm r}_i|$, 
near the $i$-th point source $r=0$ , the four harmonic functions $H$, $K$, $L$ and $M$ behave as
\begin{eqnarray}
&&H\simeq  \frac{n_i}{r}+\sum_{j(\not=i)}\frac{n_j}{|z_{ji}|},\qquad  K\simeq\frac{k_i}{r}+\sum_{j(\not=i)}\frac{k_j}{|z_{ji}|}, \notag \\
&&L\simeq\frac{l_i}{r}+l_0+\sum_{j(\not=i)}\frac{l_j}{|z_{ji}|},\qquad  M\simeq m_0+\frac{m_i}{r}+\sum_{j(\not=i)}\frac{m_j}{|z_{ji}|},
\end{eqnarray}
and the functions $f^{-1}$ and $\omega_\psi$ are approximated by
\begin{eqnarray}
f^{-1}\simeq \frac{k_i^2/n_i+l_i }{r}+c_{1(i)}',\quad \omega_\psi &\simeq&\frac{k_i^3/n_i^2+3k_il_i/2n_i+m_i}{r}+{c}_{2(i)}'. 
\end{eqnarray}
Here,  the constants $c_{1(i)}'$ and $c_{2(i)}'$ are defined  by
\begin{eqnarray}
c_{1(i)}'&:=&l_0+\sum_{j(\not=i)}\frac{1}{n_i^2|z_{ji}|}[2n_ik_ik_j-k_i^2n_j+n_i^2l_j],\\
c_{2(i)}'&:=&m_0+\frac{3}{2h_i}k_il_0\nonumber\\
&&+\sum_{j(\not =i)}\frac{1}{2n_i^3|z_{ji}|}[-(4k_i^3+3n_ik_il_i)n_j+3n_i(2k_i^2+n_il_i)k_j+3n_i^2k_il_j+2n_i^3m_j].
\end{eqnarray}
The asymptotic behaviors of the $1$-forms $\tilde\omega_i$ and $\hat\omega_{ij}$ are
\begin{eqnarray}
&&\tilde\omega_i\simeq \cos\theta d\phi,\qquad \tilde\omega_j\simeq -\frac{z_{ji}}{|z_{ji}|} d\phi \ (j\not=i),\\
&&\hat\omega_{ij}\simeq -\frac{\cos\theta}{|z_{ji}|} d\phi \ (j\not=i),\qquad \hat\omega_{jk}\simeq \frac{z_{ji}z_{ki}}{|z_{ji}z_{ki}|z_{kj}} d\phi \ (j,k\not=i,j\not = k), 
\end{eqnarray}
which leads to
\begin{eqnarray}
\hat\omega
&=&  \biggl[\sum_{j(\not=i)}\left(n_im_j-n_jm_i+\frac{3}{2}k_i l_j- \frac{3}{2}k_jl_i\right)\frac{-\cos\theta}{|z_{ji}|} + \sum_{j,k\not=i(j\not=k)}\left(n_jm_k+\frac{3}{2}k_j l_k \right) \frac{z_{ji}z_{ki}}{|z_{ji}z_{ki}|z_{kj}}  \nonumber\\
&&-\left(m_0n_i+\frac{3}{2}l_0k_i\right)\cos\theta -\sum_{j(\not=i)}\left(-m_0+\frac{3}{2}l_0k_j\right) \frac{-z_{ji}}{|z_{ji}|}+c \biggr]d\phi,
\end{eqnarray}
and
\begin{eqnarray}
\chi=  h_i \hat\omega_i+\sum_{j(\not=i)} h_j \hat\omega_j\simeq \left(n_i\cos\theta-\sum_{j(\not=i)}n_j\frac{z_{ji}}{|z_{ji}|} \right)d\phi.
\end{eqnarray}
It is obvious that the metric components $g_{rr}$ and $g_{r\psi'}$ apparently divers at $r=0$. 
However, the apparent divergence can be eliminated  by introducing new coordinates $(v,\psi')$ given by
\begin{eqnarray}
dv=dt-\left(\frac{A_0}{r^2}+\frac{A_1}{r}\right)dr,\qquad 
d\psi'=d\psi-\sum_{j(\not=i)}n_j\frac{z_{ji}}{|z_{ji}|}d\phi-\frac{B_0}{r}dr,
\end{eqnarray}
where the constants $A_0$ and $B_0$ are determined by demanding the $1/r^2$ term in $g_{rr}$ and the $1/r$ term $g_{r\psi'}$ should vanish and the constant $A_1$ is determined to remove the $1/r$ term in $g_{rr}$, which results in
 \begin{eqnarray}
 {A_0}&=&\frac{1}{2} \sqrt{3  {k_i}^2  {l_i}^2+4 n_i  {l_i}^3-4m_i(2k_i^3+3n_ik_il_i+n_i^2m_i)},\\
 {A_0B_0}&=&\frac{2  {k_i}^3+3  {k_i}  {l_i} n_i+2m_in_i^2 }{2},\\
 {4A_0A_1}&=&-4k_i^3m_0+3l_ik_i^2-6n_ik_il_im_0+6n_il_i^2-(4m_0n_i^2+6k_in_i)m_i\nonumber\\
 &+&\sum_{j(\not=i)}\frac{1}{|z_{ji}|}
 [2(l_i^3-3k_il_im_i-2n_im_i^2)n_j
 +3(k_il_i^2-4k_i^2m_i-2n_il_im_i)k_j\nonumber\\
 &&+3(k_i^2 l_i + 2n_i l_i^2-2n_ik_im_i )l_j
 -2(2 k_i^3 + 3n_i k_i l_i+2n_i^2m_i)m_j].
 \end{eqnarray}
In terms of this coordinate system, we see that the metric is then analytic in $r $ and therefore can be uniquely extended into the $r<0$ region. Moreover, one can easily confirm that the null surface $r=0$ is the Killing horizon for the Killing field $V=\partial/\partial v$. 

Taking the limit as  $(v,r)\to (v/\epsilon,\epsilon r)$ and $\epsilon \to 0$ \cite{Reall:2002bh}, after short computations, we obtain  the near-horizon geometry as
\begin{eqnarray}
ds^2_{\rm NH}&=&\frac{R_{2(i)}^2}{4}\left[d\psi'+n_i\cos\theta d\phi -\frac{2k_i(2k_i^2+3n_il_i)-4n_i^2m_i}{R_{1(i)}^4R_{2(i)}^2}rdv\right]^2+R_{1(i)}^2(d\theta^2+\sin^2\theta d\phi^2)\nonumber \\
&&-\frac{4r^2}{R_{1(i)}^2R_{2(i)}^2}dv^2-\frac{4}{R_{2(i)}}dvdr,
\label{NHmetric}
\end{eqnarray}
and 
\begin{eqnarray}
A=\frac{\sqrt{3}}{2}\left[\frac{n_ir}{k_{i}^2+n_il_{i}}dv+\frac{2k_{i}^3+3n_ik_{i}l_{i}+2n_i^2m_i}{2n_iR_{1(i)}^2}(d\psi'+n_i\cos\theta d\phi)\right], 
\label{NHgauge}
\end{eqnarray}
where we have defined 
\begin{eqnarray}
&&R_{1(i)}^2:=k_i^2+n_il_i,\label{sec:R1ineq}\\
&&R_{2(i)}^2:=\frac{3  {k_i}^2  {l_i}^2+4 n_i  {l_i}^3-4m_i(2k_i^3+3n_ik_il_i+n_i^2m_i)}{R_{1,i}^4}.\label{sec:R2ineq}
\end{eqnarray}
This is isometric to the near-horizon geometry of the BMPV black hole. 
In order to remove CTCs near all horizons, we must require the inequalities
\begin{eqnarray}
R_{1(i)}^2>0,\quad R_{2(i)}^2>0. \label{eq:CTC}
\end{eqnarray}
As will be shown, it can be expected that these are also sufficient conditions for the avoidance of CTCs throughout the outside region of black holes. 
The cross section of the event horizon can be extracted by $v={\rm const.}$ and $r=0$ in (\ref{NHmetric}) as
\begin{eqnarray}
ds^2_{\rm H}=\frac{R_{2(i)}^2}4 (d \psi'+n_i \cos\theta d\phi)^2+R_{1(i)}^2(d\theta^2+\sin^2\theta d\phi^2) \,, 
\end{eqnarray}
which is the squashed metric of the lens space $L(n_i,1)=S^3/\mathbb Z_{n_i}$.

\subsection{Axis}
We demand that there should exist no Dirac-Misner string throughout the spacetime. 
It is sufficient to impose $\hat\omega_\phi=0$ on the $z$-axis of ${\mathbb E}^3$ (i.e., $x=y=0$) in the Gibbons-Hawking space.  For the black lens solution with a single horizon~\cite{Kunduri:2014kja,Tomizawa:2016kjh}, the absence of the Dirac-Misner string among each bubble is a direct consequence of the bubble equations, whereas for the multi-black lens solution obtained here, this is not the case.

The $z$-axis of ${\mathbb E}^3$ in the Gibbons-Hawking space splits up into  the $(n+1)$ intervals: $I_-=\{(x,y,z)|x=y=0,  z<z_1\}$, $I_i=\{(x,y,z)|x=y=0,z_i<z<z_{i+1}\}\ (i=1,...,n-1)$ and $I_+=\{(x,y,z)|x=y=0,z>z_n\}$. On the $z$-axis, the $1$-forms $\hat\omega_{ij}$ and $\tilde\omega_i$  take simple forms, respectively, 
\begin{eqnarray}
\hat \omega_{ij}=\frac{(z-z_i)(z-z_j)}{z_{ji}|z-z_i||z-z_j|}d\phi,\qquad \tilde\omega_i=\frac{z-z_i}{|z-z_i|}d\phi.
\end{eqnarray}
In particular, on $ I_\pm$, $\hat\omega_{ij}$ and $\tilde\omega_i$ become, respectively,
\begin{eqnarray}
\hat \omega_{ij}=\frac{1}{z_{ji}}d\phi,\qquad \tilde\omega_i=\pm d\phi.
\end{eqnarray}
Hence,  on $I_\pm$, $\hat\omega$ automatically vanishes since
\begin{eqnarray}
\hat\omega&=&\sum_{k,j(k\not=j)}\left(h_km_j+\frac{3}{2}k_kl_j\right)\hat\omega_{kj}-\sum_{j}\left(m_0h_j+\frac{3}{2}k_j\right)\hat\omega_j+c d\phi\notag\\
&=&\sum_{k,j(k\not=j)}\left(h_km_j+\frac{3}{2}k_kl_j\right)\frac{d\phi}{z_{jk}}\mp\sum_{j}\left(m_0h_j+\frac{3}{2}k_j\right)d\phi -\sum_{k,j(k\not=j)}\left(h_km_j+\frac{3}{2}k_kl_j\right)\frac{d\phi}{z_{jk}} \notag\\
&=&\mp\sum_{j}\left(m_0h_j+\frac{3}{2}k_j\right)d\phi \notag\\
&=&\mp\left( m_0\sum_jn_j+\frac{3}{2}\sum_jk_j \right) d\phi\notag\\
&=&0,
\end{eqnarray}
where we have used Eqs.~(\ref{eq:sum-n_i}) and (\ref{eq:m0}) in the last equality.

\medskip
On the intervals $I_i\ (i=1,\cdots,n-1)$, we should impose  $\hat\omega_\phi=0$ since it does not automatically vanish there.
Let us note that if and only if  the constants $\hat\omega_\phi[I_1]-\hat\omega_\phi[I_-],\ \hat\omega_\phi[I_i]-\hat\omega_\phi[I_{i-1}]\ (i=2,\cdots,n-1), \hat\omega_\phi[I_+]-\hat\omega_\phi[I_{{n-1}}]$
vanish, the $1$-form $\hat\omega$ vanishes on all the intervals  $I_i$ $(i=1,\cdots,n-1)$.

\medskip
The constant $\hat\omega_\phi[I_1]-\hat\omega_\phi[I_-]$ can be written as
\begin{eqnarray}
\hat\omega_\phi[I_1]-\hat\omega_\phi[I_-]&=&-\sum_{j\not=1}\left[h_1m_j-h_jm_1+\frac{3}{2}(k_1l_j-k_jl_1)\right]\frac{1}{z_{j1}}+\sum_{k,j(k\not=j,k,j\not=1)}\left(h_km_j+\frac{3}{2}k_kl_j\right)\frac{1}{z_{jk}}\nonumber\\
&&-\left(m_0h_1+\frac{3}{2}k_1\right)+\sum_{j\not=1}\left(m_0h_j+\frac{3}{2}k_j\right)+c\nonumber\\
&&-\sum_{j\not=1}\left[h_1m_j-h_jm_1+\frac{3}{2}(k_1l_j-k_jl_1)\right]\frac{1}{z_{j1}}-\sum_{k,j(k\not=j,k,j\not=1)}\left(h_km_j+\frac{3}{2}k_kl_j\right)\frac{1}{z_{jk}}\nonumber\\
&&-\sum_{j}\left(m_0h_1+\frac{3}{2}k_j\right)-c\nonumber\\
&=&-2\sum_{j\not=1}\left[h_1m_j-h_jm_1+\frac{3}{2}(k_1l_j-k_jl_1)\right]\frac{1}{z_{j1}}-2\left(m_0h_1+\frac{3}{2}k_1\right). \label{eq:omega1}
\end{eqnarray}
Similarly, the constant $\hat\omega_\phi[I_i]-\hat\omega_\phi[I_{i-1}]$ can be simplified as
\begin{eqnarray}
\hat\omega_\phi[I_i]&-&\hat\omega_\phi[I_{i-1}]\nonumber \\
&=&\sum_{k,j\le i(k\not=j)}\left(h_km_j+\frac{3}{2}k_kl_j\right)\frac{1}{z_{jk}}-\sum_{k,j(j\le i <k)}\left(h_km_j-h_jm_k+\frac{3}{2} (k_kl_j-k_jl_k)\right)\frac{1}{z_{jk}}\nonumber\\
&&+\sum_{k,j(k,j>i,k\not=j)}\left(h_km_j+\frac{3}{2}k_kl_j\right)\frac{1}{z_{jk}}-\sum_{j(\le i)}\left(m_0h_j+\frac{3}{2}k_j\right)+\sum_{j(>i)}\left(m_0h_j+\frac{3}{2}k_j\right)+c\nonumber\\
&&-\sum_{k,j\le i-1(k\not=j)}\left(h_km_j+\frac{3}{2}k_kl_j\right)\frac{1}{z_{jk}}+\sum_{k,j(j\le i-1 <k)}\left(h_km_j-h_jm_k+\frac{3}{2} (k_kl_j-k_jl_k)\right)\frac{1}{z_{jk}}\nonumber\\
&&-\sum_{k,j(k,j>i-1,k\not=j)}\left(h_km_j+\frac{3}{2}k_kl_j\right)+\sum_{j(\le i-1)}\left(m_0h_j+\frac{3}{2}k_j\right)\nonumber\\
&&-\sum_{j(>i-1)}\left(m_0h_j+\frac{3}{2}k_j\right)-c\nonumber\\
&=&\sum_{j(< i)}\left(h_im_j-h_jm_i+\frac{3}{2}(k_il_j-k_jl_i)\right)\frac{1}{z_{ji}}+\sum_{j(< i)}\left(h_im_j-h_jm_i+\frac{3}{2}(k_il_j-k_jl_i)\right)\frac{1}{z_{ji}}\nonumber\\
&&-\sum_{k(>i)}\left(h_km_i-h_im_k+\frac{3}{2}(k_kl_i-k_il_k)\right)\frac{1}{z_{ik}}-\sum_{j(>i)}\left(h_im_j-h_jm_i+\frac{3}{2}(k_il_j-k_jl_i)\right)\frac{1}{z_{ji}}\nonumber\\
&&-\left(m_0h_i+\frac{3}{2}k_i\right)-\left(m_0h_i+\frac{3}{2}k_i\right)\nonumber\\
&=&2\sum_{j(<i)}\left(h_im_j-h_jm_i+\frac{3}{2}(k_il_j-k_jl_i)\right)\frac{1}{z_{ji}}-2\sum_{j(>i)}\left(h_im_j-h_jm_i+\frac{3}{2}(k_il_j-k_jl_i)\right)\frac{1}{z_{ji}}\nonumber\\
&&-2\left(m_0h_i+\frac{3}{2}k_i\right).\label{eq:omega3}
\end{eqnarray}
The constant $\hat\omega_\phi[I_+]-\hat\omega_\phi[I_{n-1}]$ becomes
\begin{eqnarray}
\hat\omega_\phi[I_+]-\hat\omega_\phi[I_{n-1}]&=&-2\sum_{j(<n)}\left(h_nm_j-h_jm_n+\frac{3}{2}(k_nl_j-k_jl_n)\right)\frac{1}{z_{nj}}\nonumber\\
&&-2\left(m_0h_n+\frac{3}{2}k_n\right),\label{eq:omega3}
\end{eqnarray}
 where we have used Eqs.~(\ref{eq:sum-n_i}) and (\ref{eq:m0}) in the last equality. 

\medskip
From Eqs.~(\ref{eq:omega1})-(\ref{eq:omega3}),  it turns out that to assure $\hat\omega=0$ on the $z$-axis,  for $i=1,\cdots,n-1$, the parameters should be subject to the constraints
\begin{eqnarray}
m_0n_i+\frac{3}{2}k_i=-\sum_{j(\not=i)}\left[n_im_j-n_jm_i+\frac{3}{2}(k_il_j-k_jl_i)\right]\frac{1}{|z_{ji}|}.\label{eq:Misner}
\end{eqnarray}

\medskip
For the discussion of the issue of orbifold singularities, we consider the rod digram of the multi-black lenses. 
On $I_\pm$, we get 
\begin{eqnarray}
\chi&=&\pm d\phi,
\end{eqnarray} 
and on $I_i$,
\begin{eqnarray}
\chi
      &=&\left(\sum_{j\le i}h_j\frac{z-z_j}{|z-z_j|}+\sum_{i+1\le j\le n}h_j\frac{z-z_j}{|z-z_j|}\right)d\phi \notag \\
      &=&\left(\sum_{j\le i}h_j-\sum_{i+1\le j\le n}h_j\right)d\phi \notag \\
      &=&\left(\sum_{j\le i}n_j-\sum_{i+1\le j\le n}n_j\right)d\phi \notag \\
      &=&\left(2\sum_{j\le i}n_j-1\right)d\phi.
\end{eqnarray}
Therefore, we can write the two-dimensional $(\phi,\psi)$-part of the metric on the intervals $I_\pm$ and $I_i$  in the form
\begin{eqnarray}
ds^2_2=(-f^2\omega_\psi^2+f^{-1}H^{-1})(d\psi+\chi_\phi d\phi)^2. \label{eq:axis}
\end{eqnarray}
Let us now use  the coordinate basis vectors $(\partial_{\phi_1},\partial_{\phi_2})$ of $2\pi$ periodicity, instead of $(\partial_\phi,\partial_\psi)$, where these coordinates are defined by $\phi_1:=(\psi+\phi)/2$ and $\phi_2:=(\psi-\phi)/2$.  
From~(\ref{eq:axis}), one can observe that  the Killing vector $v:=\partial_\phi-\chi_\phi\partial_\psi$ vanishes on each interval. 
Namely, 
\begin{enumerate}
\item on the interval $I_+$, the Killing vector $v_+:=\partial_\phi-\partial_\psi=(0,-1)$ vanishes,
\item on each interval $I_i$ ($i=1,...,n-1$), the Killing vector $v_i:=\partial_\phi-(2n-2i+1)\partial_\psi=(1-\sum_{j(\le i)}n_j,-\sum_{j(\le i)}n_j)$ vanishes,
\item on the interval $I_-$, the Killing vector $v_-:=\partial_\phi+\partial_\psi=(1,0)$ vanishes. 
\end{enumerate}
From these, we can observe that the Killing vectors $v_\pm,\ v_i$ on the intervals satisfy 
\begin{eqnarray}
{\rm det}\ (v_+^T,v_{n-1}^T)=n_n,\qquad {\rm det}\ (v_{i}^T,v_{i-1}^T)=n_i, \label{lens_cond1}
\label{noorbifold}
\end{eqnarray}
with 
\begin{eqnarray}
{\rm det}\ (v_1^T,v_{-}^T)=n_1.
\label{lens_cond2}
\end{eqnarray}
Eq. (\ref{noorbifold}) assures that the metric smoothly joints at the end points $z=z_i\ (1\le i\le n)$ of the intervals~\cite{Hollands:2007aj}, which means that there exist no orbifold singularities at adjacent 
intervals. 
Furthermore, Eqs. (\ref{lens_cond1}) and (\ref{lens_cond2}) show that the spatial cross section of the $i$-th Killing horizon ${\bm r}={\bm r}_i$ is topologically the lens space $L(n_i,1)=S^3/{\mathbb Z}_{n_i}$.

\section{Physical properties}
\label{sec:analysis}
In this section, we study physical properties of the multi-black lenses.  
\subsection{Conserved quantities}

Let us investigate conserved quantities of the multi-black lens solution. 
The ADM mass and two ADM angular momenta  can be computed as
\begin{eqnarray}
M&=&\frac{\sqrt{3}}{2}|Q|=3\pi\left[\left(\sum_{i}k_i\right)^2+\sum_{i}l_i\right],\\
J_\psi&=&4\pi \left[ \left(\sum_{i}k_i\right)^3
+\sum_{i}m_i+\frac{3}{2}\left(\sum_{i}k_i\right)\left(\sum_il_i\right)\right],\\
J_\phi&=&6\pi\left[  -\left(\sum_{i}k_i\right) \left(
\sum_{i}z_i\right)+\left(\sum_{i }k_iz_i\right)  \right], 
\end{eqnarray}
where $Q$ is a electric charge, which saturates Bogomol'ny bound. 

\medskip
The surface gravity and the angular velocities of the horizon vanish, as expected
for supersymmetric black objects in the asymptotically flat spacetime~\cite{Gauntlett:1998fz}.
The area of the $i$-horizon reads from (\ref{NHmetric}) as
\begin{eqnarray}
{\rm Area}=8\pi^2 R_{1(i)}^2 R_{2(i)} . 
\end{eqnarray}

The interval $I_i$ ($i=1,...,n-1$) represents  the bubble between adjacent two horizons which is topologically an annulus $S^1\times [0,1]$. 
The magnetic flux through $I_i$ is defined as 
\begin{eqnarray}
q[I_i]:=\frac{1}{4\pi}\int_{I_i}F\,.
\end{eqnarray}
Since the Maxwell gauge potential $1$-form $A_\mu$ is smooth at the horizons and bubbles, these fluxes are given by only the contribution from the horizons $q[I_i]=[-A_\psi]^{z=z_{i+1}}_{z=z_i}$, which leads to
\begin{eqnarray}
q[I_i]=\frac{\sqrt{3}}{2}\left[\frac{k_il_i+2n_im_i}{2(k_i^2+n_il_i)}-\frac{k_{i+1}l_{i+1}+2n_{i+1}m_{i+1}}{2(k_{i+1}^2+n_{i+1}l_{i+1})}\right] \quad (i=1,... n-1).
\label{mag_fluxes}
\end{eqnarray}

\medskip
Let us see whether there exists the parameter region such that magnetic fluxes vanish. 
For simplicity, we now consider the two-black lens solution $(n=2)$. 
From Eq.~(\ref{mag_fluxes}), the magnetic flux $q[I_1]$ between the two horizons can be written as
\begin{eqnarray}
q[I_1]=\frac{\sqrt{3}}{2}\left(\frac{k_1l_1+2n_1m_1}{2(k_1^2+n_1l_1)}-\frac{k_2l_2+2n_2m_2}{2(k_2^2+n_2l_2)}\right).\label{mag_fluxes_2}
\end{eqnarray}
From Eq.~(\ref{mag_fluxes_2}), when $m_2$ is denoted by
\begin{eqnarray}
m_2=\frac{k_1l_1(k_2^2+n_2l_2)-k_2l_2(k_1^2+n_1n_1)}{2n_2(k_1^2+n_1l_1)},
\end{eqnarray}
the magnetic flux $q[I_1]$ vanishes.
Here, let us recall that for this case, the condition~(\ref{eq:Misner}) for the absence of Dirac-Misner string singularities on the $z$-axis is simply written as 
\begin{eqnarray}
-\frac{3}{2}n_1(k_1+k_2)+\frac{3}{2}k_1=-\left[(n_1m_2-n_2m_1)+\frac{3}{2}(k_1l_2-k_2l_1)\right]\frac{1}{z_{21}},\label{eq:misner_2}
\end{eqnarray}
which gives
\begin{eqnarray}
z_{21}=\frac{k_1l_2-k_2l_1+2n_1m_2/3}{n_1k_2-n_2k_1},
\end{eqnarray} 
where we have put $m_1=0$ from Eq.~(\ref{eq:trans}) without loss of generality.
From our assumption, the constant $z_{21}$ must be positive. As shown in~FIG.~1 there exists a parameter region such that the magnetic flux vanishes for $R_1^2>0$, $R_2^2>0$ and $z_{21}>0$.

\begin{figure}[t]
\label{fig:magnetic0}
\begin{center}
\includegraphics[width=7cm]{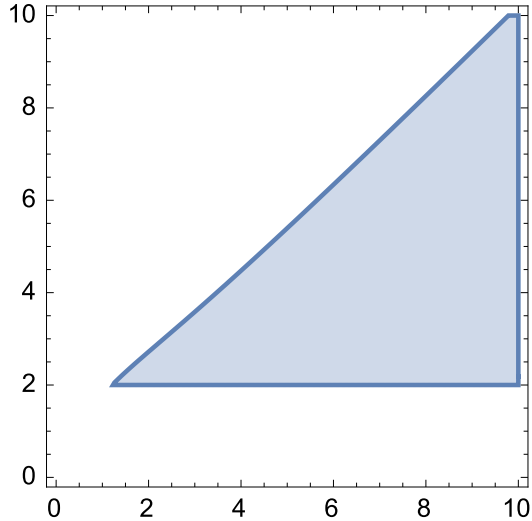}
\caption{
Region in a $(k_1,k_2)$-plane such that $R_1^2>0,\ R_2^2>0,\ z_{21}>0$ and $q[I_1]=0$  for $n_1=5$ $n_2=-4$, $l_1=l_2=1$.
}
\end{center}
\end{figure}

\subsection{No CTCs}

We demand that the domain of outer communication in the five-dimensional spacetime does not admits CTCs. 
This is achieved if the inequalities 
\begin{eqnarray}
g_{\theta\theta}>0, \qquad g_{\psi\psi}>0 \,, \qquad 
g_{\psi\psi}g_{\phi\phi}-g_{\psi\phi}^2>0 \,. 
\end{eqnarray}
are satisfied on and outside all horizons. 
Explicitly, these conditions are replaced by 
\begin{align}
\label{}
D_1:& = K^2+HL >0 \,, \\ 
D_2:& = \frac 34 K^2L^2-2K^3 M-3 HKLM+HL^3-H^2 M^2>0 \,, \\
D_3:& =  D_2 r^2\sin^2\theta-\hat \omega_\phi^2>0 \,.
\end{align} 
It is a considerably troublesome problem to prove their positivity.

As seen in FIG.~2, 
for $n=2$, we have checked the absence of CTCs by seeing numerically the positivity of $D_i\ (i=1,2,3)$ and found that there appear no causal violations in the domain of outer communications. We can expect that for $n=2$, the inequalities (\ref{eq:CTC}) are sufficient to remove CTCs in the whole domain of outer communication. 
We also expect that even for $n>2$,  (\ref{eq:CTC}) are sufficient to remove causal pathologies on and outside the horizon.

\begin{figure}[t]
\label{fig:CTC}
\begin{center}
\includegraphics[width=15cm]{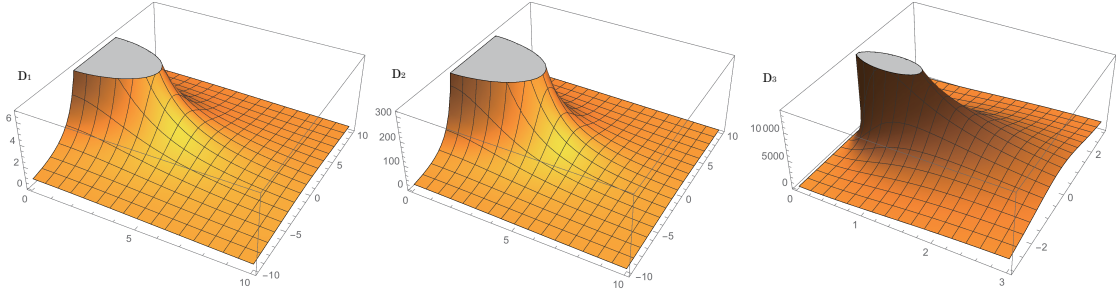}
\caption{Plots of $D_i$'s against ($\rho=\sqrt{x^2+y^2}, z$) 
for $n_1=2$ $n_2=-1$ , $z_1=0$, $z_2\simeq 0.0011$, $k_1=k_2=4$, $l_1=l_2=1$, $m_1=0$, $m_2=0.001$. 
No naked CTC appears.}
\end{center}
\end{figure}

\subsection{Critical surfaces}
 One of physically interesting features is that  the harmonic function $H$ becomes negative around ${\bm r}={\bm r}_i$ ($i=1,\cdots,n$), which leads to the ($-,-,-,-$) signature of the Gibbons-Hawking base space. 
However, the signature of the five-dimensional spacetime metric remains Lorenzian because the function $f^{-1}H$ is positive. 
In this case, a so-called evanescent ergosurface appears \cite{Bena:2007kg} at the places which $f=0$ corresponding to $H=0$.

At this surface, one must impose the regularity condition $K \ne 0$ when $H=0$ since if this does not hold, the spacetime could not become regular there.  
In other words, if there exist points $z=z_c$ on the axis 
such that 
\begin{eqnarray}
H=\sum_{i}\frac{n_i}{|z_c-z_i|}=0,\qquad 
K=\sum_{i}\frac{k_i}{|z_c-z_i|}=0, 
\label{HKsingCS}
\end{eqnarray}
these critical surfaces are singular, which leads to
\begin{eqnarray}
\sum_{i\ge 2}\frac{n_1k_i-n_ik_1 }{|z_c-z_i|}=0. 
\end{eqnarray}
Hence, for instance, one of the sufficient conditions to avoid the singularities at critical surfaces 
is that for any $i$ $(i=1,\cdots,n)$, $n_1k_i-n_ik_1$ must have the same signs   
\begin{eqnarray}
n_1k_i-n_ik_1 > 0, \qquad{\rm or,} \qquad n_1k_i-n_ik_1< 0.
\label{eq:additional}
\end{eqnarray}

For $n=2$, 
one of evanescent ergosurfaces exists at 
\begin{eqnarray}
z=\frac{n_1z_2+n_2z_1}{n_1+n_2}
\end{eqnarray}
for $z\in I_+$ when $n_1>0$ and $z\in I_-$ when $n_1<0$, whereas the other exists at
\begin{eqnarray}
z=\frac{n_1z_2-n_2z_1}{n_1-n_2}
\end{eqnarray}
for $z\in I_1$.

\section{Summary}
\label{sec:discuss}

In this work, we have constructed an asymptotically flat and stationary multi-black lens solution as a supersymmetric solution in the bosonic sector of the five-dimensional minimal supergravity. 
We have shown that this solution describes mechanical equilibrium state of an arbitrary number of charged black lenses and the degenerate Killing horizons admit different lens space topologies $L(n_i,1)=S^3/{\mathbb Z}_{n_i}\ (i=1,\cdots,n)$, where  each $n_i$ takes non-zero different integers but must satisfy the constraint equation $\sum_{i=1}^nn_{i}=1$. 
This multi-black lens spacetime has a spatial symmetry of $U(1)\times U(1)$ because all horizons are alined on the $z$-axis in the Gibbons-Hawking space. 
Moreover, we have also computed the conserved charges including the (positive and BPS-saturating) mass, two angular momenta and the magnetic fluxes on the bubbles.

\medskip
As for the supersymmetric black lens solution in obtained in~\cite{Kunduri:2014kja,Tomizawa:2016kjh}, there exists no limit such that all the magnetic fluxes vanish. Therefore, one can consider that, at least, for the supersymmetric black lens with the single horizon of the topology $L(n,1)=S^3/{\mathbb Z}_n$ $(n=2,3,\cdots)$ in~\cite{Kunduri:2014kja,Tomizawa:2016kjh},  the existence of the magnetic fluxes plays an essential role in supporting the horizon of the black lens. On the other hand, for the supersymmetric multi-black lenses obtained in this paper, it seems that the magnetic flux does not necessarily need to exist. 

\medskip
In this work, we have considered the supersymmetric solution subject to the constraint (\ref{eq:sum-n_i}), which comes from the requirement that the topology of spatial infinity $r\to\infty$ should be $S^3$, when the spacetime asymptotically becomes flat. This constraint seems to impose a considerably strong restriction on the topologies of horizons.   However, if one replaces the harmonic function $H$ in the Gibbons-Hawking base space with, for instance, another one
\begin{eqnarray*}
\sum_{i=1}^n\frac{n_i}{r_i}-\sum_{i=n+1}^{N}\frac{1}{r_i}\  \left(N:=\sum_{i=1}^n{n_i}-1\right)
\end{eqnarray*}
and moreover, if at each point ${\bm r}={\bm r}_i\ (i=n+1,\cdots,N)$ where the harmonic function diverges, one demand regularity (this corresponds to the conditions $c_2=0$ in~\cite{Tomizawa:2016kjh}) 
one no longer may need to impose the constraint (\ref{eq:sum-n_i}). 
In this case, each horizon ${\bm r}={\bm r}_i\ (i=1,\cdots,n)$ can have an independent lens space topology of $L(n_i,1)$.

\acknowledgments
This work was partially supported by the Grant-in-Aid for Young Scientists (B) (No.~26800120) from Japan Society for the Promotion of Science (S.T.).

\appendix

\end{document}